\documentclass[12pt,twocolumn]{aastex}      

\usepackage[latin1]{inputenc}
\usepackage{color}

\shorttitle{Zooming into the Collimation Zone in a Massive Protostellar Jet}
\shortauthors{Carrasco-Gonz\'alez et al.}

\begin{document}

\title{Zooming into the Collimation Zone in a Massive Protostellar Jet}

\correspondingauthor{Carlos Carrasco-Gonz\'alez}
\email{c.carrasco@irya.unam.mx}

\author{Carlos Carrasco-Gonz\'alez}
\affiliation{Instituto de Radioastronom\'{\i}a y Astrof\'{\i}sica (IRyA-UNAM), Morelia, Mexico}

\author{Alberto Sanna}
\affiliation{INAF - Osservatorio Astronomico di Cagliari, Selargius, Italy}
\affiliation{Max-Planck-Institut f\"{u}r Radioastronomie (MPIfR), Bonn, Germany}

\author{Adriana Rodr\'{\i}guez-Kamenetzky}
\affiliation{Instituto de Radioastronom\'{\i}a y Astrof\'{\i}sica (IRyA-UNAM), Morelia, Mexico}

\author{Luca Moscadelli}
\affiliation{INAF - Osservatorio Astrofisico di Arcetri, Firenze, Italy}

\author{Melvin Hoare}
\affiliation{School of Physics \& Astronomy, University of Leeds, Leeds, UK}

\author{Jos\'e M. Torrelles}
\affiliation{Institut de Ci\`encies de l'Espai (ICE, CSIC), Barcelona, Spain}
\affiliation{Institut d'Estudis Espacials de Catalunya (IEEC), Barcelona, Spain}

\author{Roberto Galv\'an-Madrid}
\affiliation{Instituto de Radioastronom\'{\i}a y Astrof\'{\i}sica (IRyA-UNAM), Morelia, Mexico}

\author{Andr\'es F. Izquierdo}
\affiliation{European Southern Observatory, Garching, Germany}
\affiliation{Leiden Observatory, Leiden University, Leiden, The Netherlands}

\begin{abstract}
Protostellar jets have a fundamental role at the earliest evolution of protostars of all masses. In the case of low-mass ($\lesssim$8~$M_\odot$) protostars, strong observational evidence exists that the launching and collimation is due to the X- and/or disk-wind mechanisms. In these models, it is the protostar/disk system that creates all the necessary conditions to launch and collimate the jets near the protostar via strong magnetic fields. The origin of jets from more massive protostars has been investigated much less, in part because of the difficulty of resolving the collimation zone in these more distant objects. Here we present the highest angular resolution observations of a jet powered by a massive protostar, the Cep A HW2 radio jet. We imaged the radio emission at projected distances of only $\sim$20~au from the protostar, resolving the innermost 100~au of a massive protostellar jet for the first time. The morphology of the radio jet emission in this massive object is very different than what is usually observed in jets from low-mass protostars. We found that the outflowing material in HW2 has two components: a wide-angle wind launched from the protostar/disk system, and a highly collimated jet starting at 20$-$30 au from the protostar. We discuss two possible scenarios: an extension of the classical disk-wind to a massive protostar, or external collimation of a wide-angle wind. These results have important consequences for our understanding of how stars of different masses are formed.
\end{abstract}

\section{Introduction}

 Understanding how stars form requires unraveling the link between accretion and ejection phenomena in protostars. We know that protostellar jets and winds, that manifest themselves as parsec-scale molecular outflows, are necessary to remove mass and angular momentum from accretion disks in order to allow a net mass flow onto the protostars. Despite jets being discovered several decades ago, the underlying mechanism of mass ejection is still debated. One of the main reasons has been the lack of angular resolution needed to zoom into the launching and collimation zone close ($\lesssim$100~au) to the protostars (e.g., see Frank et al. 2014 for a review on protostellar jets).

 Several mechanisms have been proposed to explain the origin of protostellar jets. The most commonly invoked mechanism consists of launching of matter via magnetocentrifugal forces and plasma confinement by a helical magnetic field, the latter generated in the disk/protostar system. This mechanism is similar to that proposed to explain relativistic jets (e.g., Blandford \& Payne 1982). Furthermore, for the case of protostellar jets, the launching mechanism is named either X-wind (Shu et al. 1994) or disk-wind (Pudritz \& Norman 1983, 1986), depending on the distance from the star where the jet material is channeled and accelerated. Both mechanisms predict that the jet is launched and highly collimated very near the protostar, at distances of $\sim$0.05~au in the X-wind scenario (e.g., Shang et al. 2007), or a few au in the disk-wind scenario (e.g., Pudritz et al. 2007). These models explain the flow collimation self-consistently, since it is the protostar/disk system itself that creates all the necessary conditions to launch and collimate the jet. Interferometric observations at centimeter wavelengths of the nearest (150-500 pc) low-mass star-forming regions allow exploring the process of collimation by resolving free-free emission at scales of a few au. In this way, it has been established that collimation of low-mass protostellar jets takes place well below some tens of au (see Anglada et al. 2018, and references therein). The highest spatial resolution observation to date of a low-mass protostellar jet, the Class I/II protostar HL Tau, suggests that the jet is already collimated below 1.5\,au from the protostar (Carrasco-Gonz\'alez et al. 2019). Moreover, in some objects, strong helical magnetic fields have been observed close to the protostar (Ray et al. 1997) or at scales of a few hundreds of au (Lee et al. 2018). Therefore, the evidence for low-mass protostellar jets strongly supports a scenario where self-collimation happens due to X- or disk-winds.
 
 Large-scale molecular outflows are also commonly associated with the earliest stages in the formation of massive (B- and O-type) stars. However, bipolar outflows observed in high-mass star-forming regions tend to be less collimated than those observed in low-mass regions (e.g., Arce et al. 2007). Recent radio observations have established that very young massive protostars are also commonly associated with compact free-free emission, being the signpost of mass ejected near the protostar (e.g., Moscadelli et al. 2016; Purser et al. 2016, 2021; Rosero et al. 2016, 2019; Sanna et al. 2018; Obonyo et al. 2019). However, it is not yet clear whether this emission traces the bases of jets as well collimated as those observed in low-mass star-forming regions. To date, we know of only a handful of highly collimated jets from massive protostars extending over tens of thousands of au (e.g., Cep A HW 2: Curiel et al. 2006; HH 80-81: Rodr\'{\i}guez-Kamenetzky et al. 2017; IRAS 16547$-$4247: Rodr\'{\i}guez et al. 2005; G035.02+0.35: Sanna et al. 2019a; IRAS 16562$-$3959: Guzm\'an et al. 2010; G016.59$-$0.05: Moscadelli et al. 2019), and, although evidence exists for magnetic fields playing an important role in their collimation (e.g., Carrasco-Gonz\'alez et al. 2010, Surcis et al. 2014, Sanna et al. 2015, Maud et al. 2019), it is not yet clear whether collimation takes place at small distances from massive protostars. One possibility is that different mechanisms could be at work to produce jets around massive protostars, as they are expected to accrete at higher rates from much larger, massive and turbulent disks than those around low-mass protostars (e.g., Carrasco-Gonz\'alez et al. 2012, Sanna et al. 2019b, A\~nez-L\'opez et al. 2020). Under these conditions, it could be very difficult to form strong and well-structured helical magnetic fields capable of highly collimate jets at au scales. Indeed, wide-angle or even spherical winds are also commonly found in high-mass star-forming regions (e.g., Torrelles et al. 1997, 2001, 2011; Moscadelli et al. 2007). Interestingly, it has been proposed that initially poorly collimated winds could be externally collimated at large ($\gtrsim$\,10--100~au) distances from the protostar by strong ambient medium pressure (e.g., Carrasco-Gonz\'alez et al. 2015) or by a large-scale ordered magnetic field (e.g., Albertazzi et al. 2014). Therefore, there is the possibility that mass ejection from massive protostars could be mostly isotropic at au scales, and, under favorable conditions, it might become collimated at larger distances from the driving source.
 
 In this Letter we present the highest angular resolution observations of a protostellar jet driven by a massive (B-type) protostar. The HW2 object, located at 700~pc (Moscadelli et al. 2009) in the Cepheus A high-mass star-forming region, is the second nearest massive protostar known to be associated with powerful outflow phenomena. This object is known to drive a large molecular outflow (Rodr\'{\i}guez et al. 1980, G\'omez et al. 1999) and has been intensively studied at radio wavelengths since the 1990s. Early Karl G. jansky Very Large Array (VLA) observations detected a relatively bright (fluxes $\sim$3-30 mJy between 1 cm and 7 mm) radio source with an elongated morphology in the same direction as the observed large-scale outflow and properties suggestive of a biconical ionized jet (Rodr\'{\i}guez et al. 1994). High angular resolution 10 yr monitoring at radio wavelengths revealed large ($\sim$500~km~s$^{-1}$) proper motions of the radio continuum knots away from the putative position of the central star, confirming the jet/wind nature of the radio emission (Curiel et al. 2006). Multiepoch water maser observations revealed that the radio jet coexists with a slower ($\sim$100~km~s$^{-1}$) wide-angle wind also powered by the HW2 object (Torrelles et al. 2011). An elongated dusty structure perpendicular to the radio jet was also detected with the SMA at 0.9~mm, which was proposed to be tracing a large ($\sim$300~au radius) and massive (1-8~$M_\odot$) disk around the protostar (Patel et al. 2005, Jim\'enez-Serra et al. 2007, Torrelles et al. 2007). Moreover, observations of CH$_3$OH masers have revealed infall motions in a plane perpendicular to the radio jet, supporting the existence of a disk accreting at a high rate of $\sim$10$^{-4}$~$M_\odot$~yr$^{-1}$ (Sanna et al. 2017). All these results make Cep A HW2 one of the best known examples of a disk/jet system associated with a massive protostar.  Our new very high angular observations allow mapping the outflowing material at distances of only $\sim$20~au from the protostar, resolving the innermost 100~au of a massive protostellar jet for the first time that provide important consequences for our understanding of how stars of different masses are formed.

\section{Data Analysis}

\subsection{Observations} \label{Observations}

 Observations were performed with the VLA of the National Radio Astronomy Observatory (NRAO)\footnote{The National Radio Astronomy Observatory is a facility of the National Science Foundation operated under cooperative agreement by Associated Universities, Inc.} at $Ka$ and $Q$ bands, by using the A configuration on 2016 October 28 and November 9 (Project Code: 16B-201). The phase center was set to (J2000) 22$^{\rm h}$56$^{\rm m}$17.971$^{\rm s}$ +62$^\circ$01$\arcmin$49.28$\arcsec$.  Phase calibration was performed by observing the strong quasar J2148+6107 every $\sim$2.5 minutes; 3C147 and 3C84 were observed as flux and bandpass calibrators, respectively. Both the $Ka$ and $Q$ bands were recorded with the standard continuum setup, consisting of 64 spectral windows of 128 MHz each. The total continuum bandwidth observed at each band was 8 GHz covering the ranges 29-37 ($Ka$ band) and 40-48 GHz ($Q$ band). The visibility data were calibrated using the Common Astronomy Software Applications (CASA) package (version 5.4.2) and the NRAO pipeline for VLA continuum observations. During the calibration process, we identified that a large part of the lowest-frequency baseband (29-33 GHz) of the $Ka$ band was noisy with a low signal-to-noise ratio. Thus, we decided to remove it and limited the frequency coverage of the Ka band data to 33-37 GHz.

 We made clean images of both bands using the task \emph{tclean} of CASA. In order to obtain the highest angular resolution possible, we apply a Briggs weighting to the uv data, setting the parameter \emph{robust=$-$2}, equivalent to uniform weighting, which yields synthesized beams of 72$\times$41~mas$^2$ with a position angle (P.A.) of $-$61~$^\circ$ for the Ka band image, and 41$\times$27~mas$^2$ with a P.A. of $-$56~$^\circ$ for the Q band image. In both images, we detect a continuum source with a similar spatial morphology, but the Ka band image appears slightly shifted by $\sim$16~mas to the NW with respect to the Q band image. This offset is consistent with the expected positional accuracy of the VLA at these frequencies. Thus, in order to combine the data at both bands to produce a single image, we corrected for this offset following two steps. First, we aligned images made with the same overlapping uv range (500-3500 k$\lambda$) by minimizing the offset positions of corresponding peaks of emission. We estimated relative offsets of $\sim$16~mas and $\sim$4~mas in right ascension and declination, respectively, which were applied to the uv data of Ka band in order to be aligned to the Q band. Second, we produced a single image with uniform weighting by cleaning simultaneously Q band visibilities with those at Ka band shifted in position. This final image is centered at a frequency of 40 GHz, corresponding to a wavelength of 7.5~mm, and attains an rms noise of 90~$\mu$Jy~beam$^{-1}$ with a synthesized beam of 42$\times$28~mas$^2$ and P.A. of $-$55~$^\circ$. The orientation of the minor axis of the beam is parallel to the well known jet direction (P.A.$\simeq$45$^\circ$), which allows us to optimally sample the emission along the jet direction with a resolution of 28~mas. The poorer resolution in the direction perpendicular to the jet does not limit our analysis, since we do not expect to resolve the jet width even with a much smaller beam. Thus, in this favorable case, it is possible to neglect the distortion of an elongated beam, justifying our final use of a circular restoring beam with a 28\,mas diameter. 

\subsection{Radiative Transfer Models} \label{RTModels}

 Several physical models for different types of ionized winds with a detailed treatment of the radiative transfer were created. For this, we first used the \emph{sf3dmodels} package\footnote{https://star-forming-regions.readthedocs.io/en/latest/} (see Izquierdo et al. 2018 for a full description) which creates three-dimensional grids of density, temperature and ionization fraction given a number of wind parameters, and builds on the toy model first proposed by Reynolds~(1986). Each wind is defined by a mass-loss rate $\dot{M}$, an injection radius $z_0$ (i.e., the distance from the protostar at which the wind starts) and initial values of the wind velocity $v_0$, ionization fraction $x_0$, temperature $T_0$ and width $\omega_0$. The width of the wind $\omega$ is assumed to vary with the distance from the protostar $z$ as $\omega(z)=\omega_0 (z/z_0)^\epsilon$. The parameter $\epsilon$ defines the wind geometry, with $\epsilon$=1 describing a conical or wide-angle wind and $\epsilon<$1 representative of a collimated jet. The initial density $n_0$ is calculated from the mass-loss rate, the initial width, and the initial velocity, and its variation along the wind direction is fully determined by the wind geometry and mass-conservation, i.e., $n(z)=n_0 (z/z_0)^{q_n}$, with $q_n=-2\epsilon$. For the temperature and ionization fraction, we also considered variations in the form of power laws, $T(z)=T_0 (z/z_0)^{q_t}$, $x(z)=x_0 (z/z_0)^{q_x}$ (cf. Reynolds~1986). For simplicity, we considered only outflowing constant velocities, $v_0$. The physical models are plugged by \emph{sf3dmodels} into RADMC-3D (Dullemond et al. 2012), which solves the radiative transfer equation for free-free emission along the different lines of sight and produces a synthetic continuum image at the desired wavelength. We simulated continuum images at a reference wavelength of 7.5~mm, which were compared to our VLA image after convolution to the same angular resolution. 

\section{Results}

\begin{figure*}[h!]
\begin{center}
\includegraphics[height=0.7\textheight]{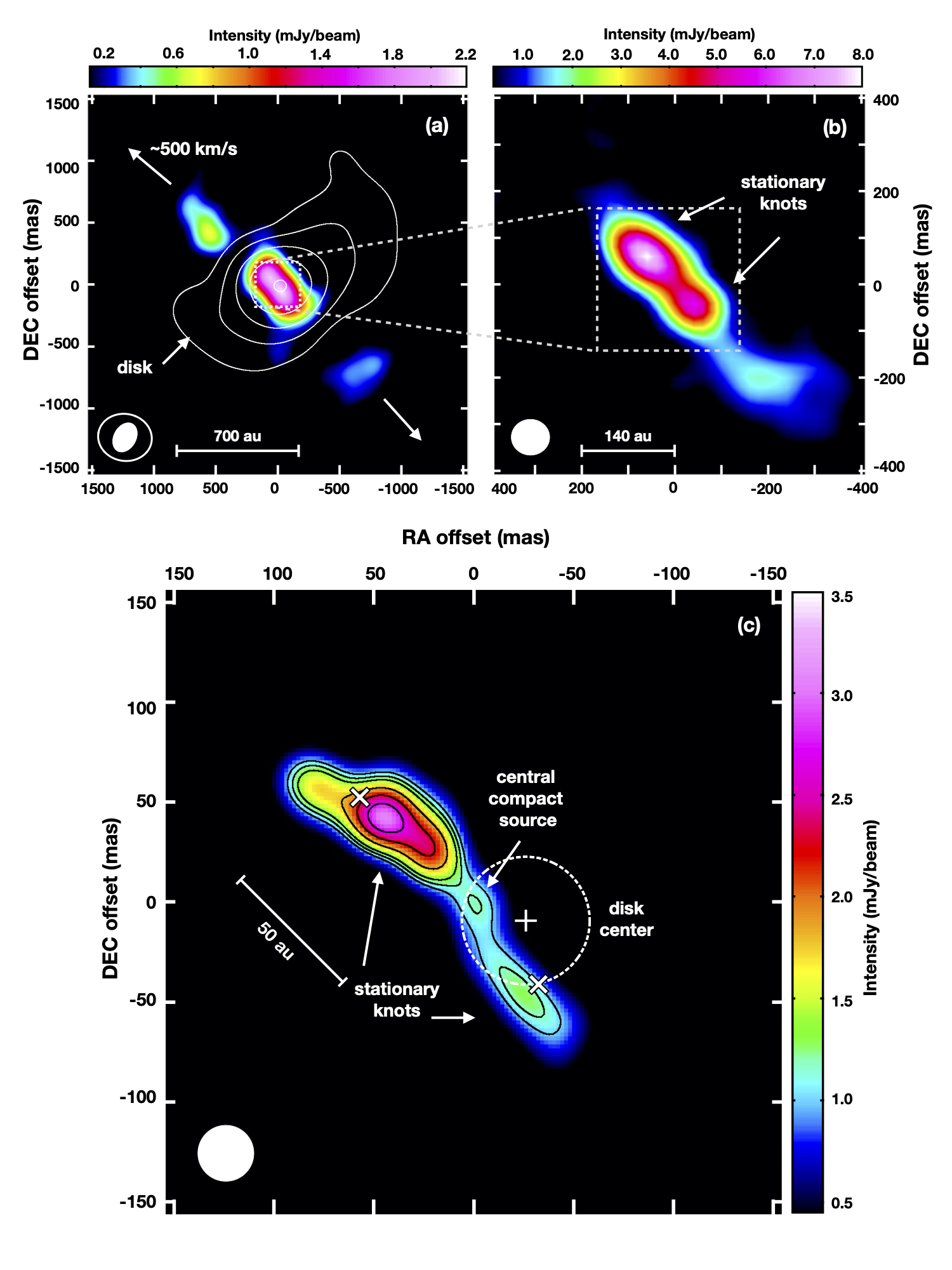}
\caption{\footnotesize{Radio continumm images of the Cep A HW2 radio jet at different resolutions. \textbf{(a)} Color scale is a 3.6 cm continuum image obtained with the VLA in 2000 (Curiel et al. 2006; angular resolution $\sim$200~mas). The emission traces ionized gas in the jet and shows a morphology consistent of an elongated core and two external knots moving away from the central core at velocities of $\sim$500~km~s$^{-1}$, which is representative of the jet velocity (Curiel et al. 2006). Contours are 1.3 mm emission recently observed with NOEMA (Beuther et al. 2018; angular resolution $\sim$400~mas) which traces dust emission from a massive disk (see Patel et al. 2005 and Jim\'enez-Serra et al. 2007). Contour values are 16, 32, 64, 128, and 256 times the rms of the map, 1.7 mJy~beam$^{-1}$. \textbf{(b)} Radio continuum image of the central core of the radio jet at 1.3 cm (from Curiel et al. 2006; angular resolution $\sim$80~mas). The elongated core seen in panel (a) is resolved here into two internal radio knots. These internal knots are stationary (Curiel et al. 2006) suggesting that they are tracing shocks of a different nature than the external moving knots. \textbf{(c)} Our new image at 7.5 mm, with a angular resolution of 28~mas, resolves the innermost region of the radio jet. The field shown in this panel corresponds to the squares shown in previous panels. Contours trace emission at 11, 13, 15, 20, 25, 30, and 40 times 90~$\mu$Jy~beam$^{-1}$. The "x" symbols mark the positions of the stationary knots seen in panel (b), which are here resolved as elongated in the direction of the jet. The cross marks the peak position of the 1.3 mm structure of panel (1) and the circle represents the absolute position error ($\sim$35 mas). A compact 7.5 mm source is detected at the center whose position is consistent with the center of the massive dusty disk. In all panels, offsets are with respect to the position of the central compact source, (J2000)~22$^{\rm h}$56$^{\rm m}$17.9854$^{\rm s}$~$+$62$^\circ$01$\arcmin$49.55$\arcsec$}}
\label{Fig1}
\end{center}
\end{figure*}

\begin{deluxetable*}{rccc}[h!]
\tablewidth{0pt}
\tablecaption{Radiative Transfer Models\label{Tab1}}
\tablehead{ 
              &   Standard Jet   &  Wide-angle Wind   &   Standard Jet \\
Parameter     &   (z$_0$=3 au)    &  (z$_0$=3 au)      &   (z$_0$=25 au) }
\startdata
Width ($\omega_0$ ; $\epsilon$)          & 10 au    ; 2/3  & 10 au    ; 1      & 10 au ; 2/3   \\
Velocity (v$_0$)                    & 500 km/s        & 100 km/s          & 500 km/s      \\
Mass-loss rate (dM/dt)              & 1.75$\times$10$^{-6}$ M$_\odot$/yr & 1.75$\times$10$^{-6}$ M$_\odot$/yr & 1.5$\times$10$^{-6}$ M$_\odot$/yr \\
Temperature (T$_0$ ; q$_T$)         & 10$^4$ K ; 0    & 10$^4$ K ; $-$0.5 & 10$^4$ K ; 0  \\
Ionization Fraction (x$_0$ ; q$_x$) & 1 ; 0           & 0.1 ; $-$0.5      & 1 ; 0         \\
\enddata
\end{deluxetable*}

\begin{figure*}[h!]
\begin{center}
\includegraphics[width=\textwidth]{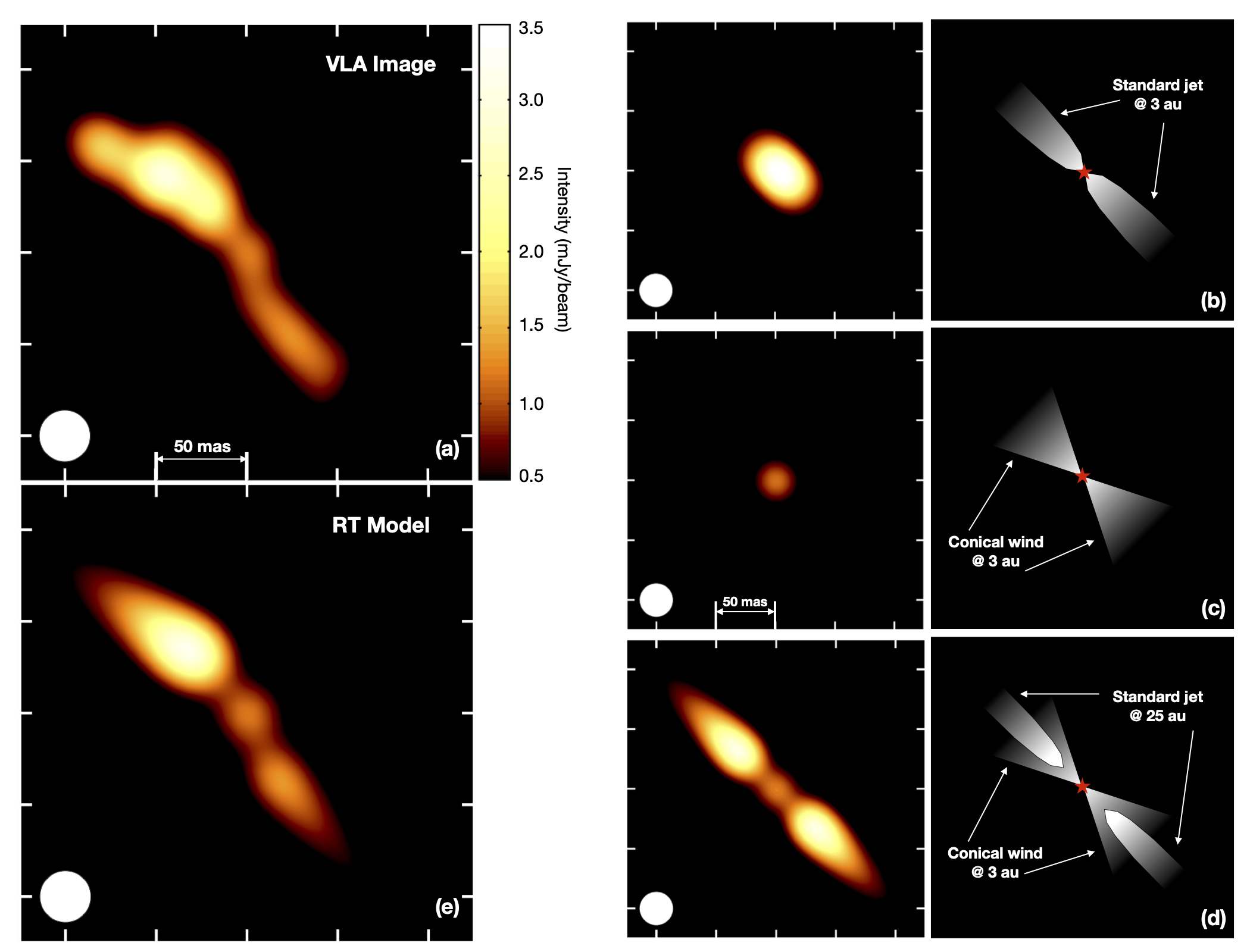}
\caption{\footnotesize{A radiative transfer modeling of the 7.5 mm image of Cep A HW2. \textbf{(a)} Image of the radio jet as shown in Figure 1c. \textbf{(b)} A model consisting of a standard jet (Reynolds 1986) starting at 3 au from the protostar. The left panel is the output image from the radiative transfer model with the same intensity scale of panel (a); the right panel shows a schematic view of the jet. As can be seen, an elongated bright source is expected. \textbf{(c)} Same as (b) for a conical wind starting at 3 au from the protostar. The output image only shows a compact source which traces the densest part of the wind. \textbf{(d)} A combination of a conical wind starting at 3 au, and a standard jet starting at 25 au from the protostar. \textbf{(e)} Same as panel (d) but the north and south parts of the jet have slightly different P.A.s and mass-loss rates.}}
\label{Fig2}
\end{center}
\end{figure*}

\begin{figure*}[h!]
\begin{center}
\includegraphics[height=0.85\textheight]{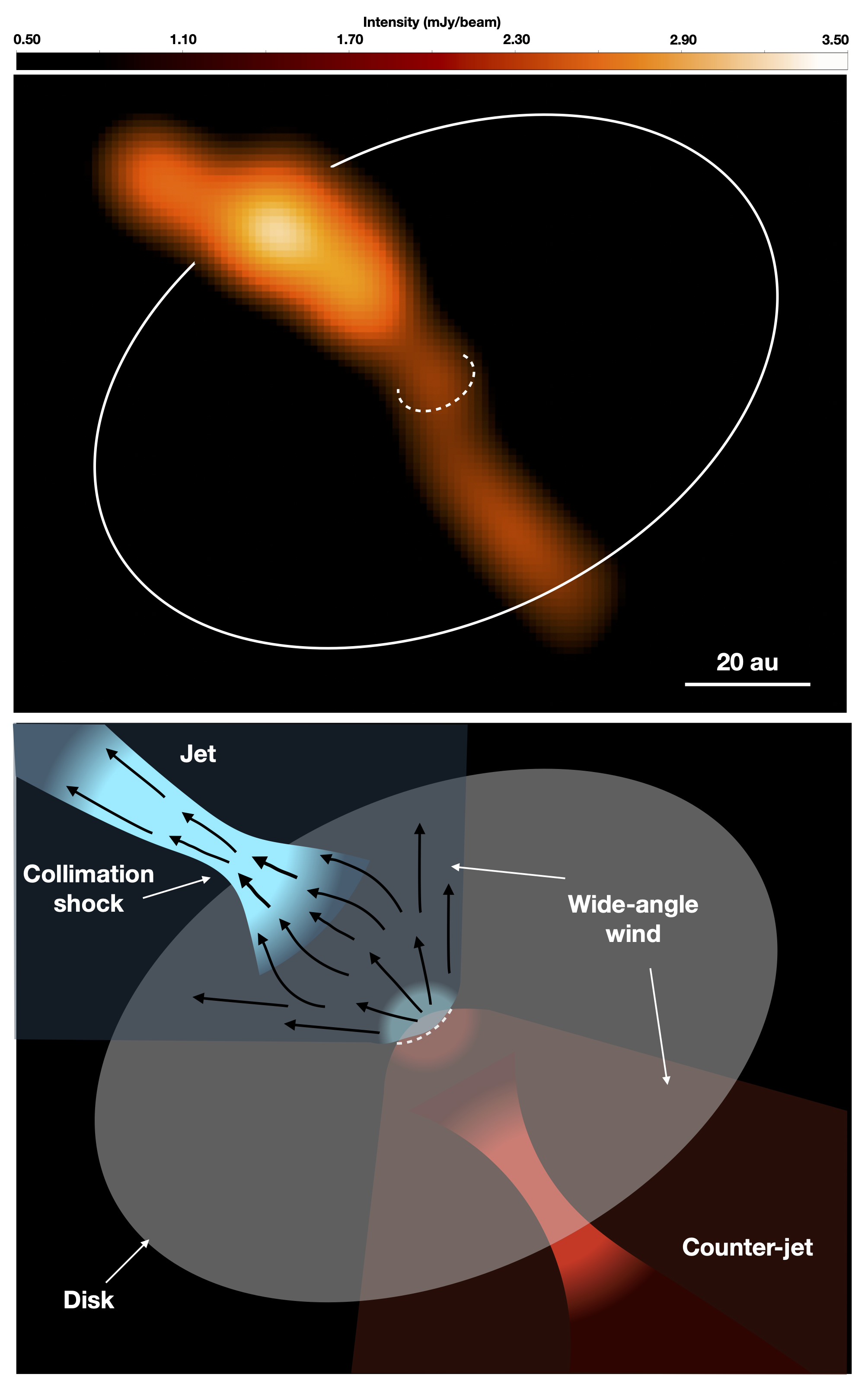}
\caption{\footnotesize{Schematic view of our interpretation of the 7.5 mm observations of the Cep A HW2 radio jet. The triple morphology of the emission implies the combination of a wide-angle wind accelerated near the protostar and a collimated jet starting at a distance of $\sim$20-30~au from the protostar. The stationary elongated knots are then tracing collimation shocks at the base of the jet and counter-jet observed at larger scales. There are two possible scenarios to explain this behaviour: a disk-wind from a massive protostar/disk system, or an externally collimated jet.}}
\label{Fig3}
\end{center}
\end{figure*}

 The available radio observations to date offer a multi-scale view of the Cep A HW2 disk/jet system, and allows progressively zooming into the launching and collimation zone of the jet (see Figure \ref{Fig1}). At scales of $\sim$1000~au, the 3.6~cm continuum emission shows a clear radio jet morphology (Figure \ref{Fig1}a), with a central, bright, elongated core, and two separated knots either side of the core, moving away in opposite directions (Curiel et al. 2006). These radio knots are commonly detected in sensitive observations of radio jets and they are associated with an increase of the electron density and/or ionization fraction due to shocks of the jet material. While radio knots trace internal shocks in the jet, due to subtle changes in the ejection velocity, the observed velocity of the knot is considered as tracing the average jet velocity (e.g., Mart\'{\i} et al. 1995). Radio knots could also be the result of strong shocks against the dense ambient medium, in which case the jet velocity should be considered to be higher (e.g., Rodr\'{\i}guez-Kamenetzky et al. 2016). Regardless of the specific scenario, the proper motions detected for these radio knots ($\sim$500~km~s$^{-1}$) imply that the material of the Cep A HW2 jet moves several times faster than in low-mass radio jets ($\sim$100~km~s$^{-1}$). Emission from the dusty disk is resolved in a direction perpendicular to the radio jet by recent NOEMA 1.3 mm high angular resolution observations (Beuther et al. 2018; see Figure \ref{Fig1}a). Higher angular resolution observations at 1.3~cm further resolved the internal $\sim$200~au of the radio jet into two radio knots separated by $\sim$0$\farcs$1 or $\sim$70~au (see Figure \ref{Fig1}b; Curiel et al. 2006). Interestingly, and contrary to the case of the external knots in Figure \ref{Fig1}a, the two internal knots did not show any significant proper motion after about ten years (Curiel et al. 2006). This evidence implies that, even though the internal stationary knots should also trace shocks in the jet, their nature might not be the same than that of the external shocks. Indeed, if the internal knots were tracing internal or terminal shocks of 100\,km\,s$^{-1}$ velocity, their position shift with time would result in measurable proper motions: at a distance of 700\,pc and over a period of 10 yr, an angular shift of about 80\,mas corresponds to a velocity of about 26\,km\,s$^{-1}$.

 Our new VLA observations at 7.5~mm, with a resolution of $\sim$28~mas, or $\sim$20~au, offers the finest view of the launching and collimation zone of the radio jet, resolving the inner $\sim$100~au from the protostar for the first time. Our image reveals an interesting morphology: two elongated knots and a compact central source (see Figure \ref{Fig1}c). The separation between the two elongated knots is $\sim$0$\farcs$1, namely, the same separation between the two stationary knots seen at lower resolution. Moreover, similarly to the lower angular resolution image, there is a difference in the intensities of the knots, with the northern one being brighter than the southern one (see Figure \ref{Fig1}b). Therefore, we conclude that the two elongated knots in our 7.5~mm image are the two stationary knots seen at lower resolution, and we also confirm the absence of proper motions 12 yrs after the previous 10-yr monitoring. The global orientation of the emission has a P.A. of $\sim$45$^\circ$, but we note slightly different P.A.s for each of the knots: $\sim$50$^\circ$ for the northern one, $\sim$40$^\circ$ for the southern one. A similar difference has been observed in the P.A.s traced by the proper motions of the external radio knots: the motions of the northern external knot trace a P.A. of $\sim$50$^\circ$ while the southern external knot moves in a direction with a P.A. of $\sim$42$^\circ$ (Curiel et al. 2006). We interpret these findings as strong evidence that the two stationary elongated knots are tracing high density material at the base of the jet. This jet produces shocks farther away from the protostar, which are then detected as external moving radio knots in lower angular resolution observations. A central compact source was already marginally detected in previous VLA images at 7~mm with a resolution of $\sim$40~mas and it was proposed to trace the material closest to the protostar (Curiel et al. 2006). Our higher angular resolution image at 7.5~mm confirms its presence, and we clearly separate this source from the two elongated knots (see Figure \ref{Fig1}c). Note that the position of the peak of the 1.3~mm dust emission is also consistent with the position of the central 7.5~mm compact source (see Figure \ref{Fig1}c). This evidence strongly suggests that the accretion disk is centered on the central compact source, which would be tracing the most recently ejected gas by the protostar/disk system.

 To better understand the underlying physical structure traced by the 7.5~mm emission, we produced a number of radiative transfer models for several combinations of wind parameters (see Table \ref{Tab1} and \S\ref{RTModels}). We aim to reproduce the main observational features, i.e., the triple morphology as well as the intensities seen in our VLA image (see Figure \ref{Fig2}); it is not intended to be a best-fit model. We consider an inclination of the system of 26$^\circ$ with respect to the plane of the sky, with the NE part of the jet/wind pointing to the observer (see Sanna et al. 2017). Since protostellar jets driven by low-mass protostars appear to be launched and collimated very near the protostar/disk system, at scales from a few au or less, we first explored a fully ionized standard collimated jet ($\epsilon$=2/3) starting at a distance of 3~au from the protostar (see Fig \ref{Fig2}b). We consider a constant outflowing velocity of $v_0$=500~km~s$^{-1}$, as observed in the external shocks, and a mass-loss rate of $\dot{M}$=1.75$\times$10$^{-6}$ M$_\odot$~yr$^{-1}$. As expected, this results in a bright elongated source with a size of $\sim$60~mas (see Figure \ref{Fig2}b). Note that a collimated jet ($\epsilon<1$) will always produce a slowly decreasing electron density, which, in turn, appears in the radio image as an elongated source with its maximum intensity close to the center. In contrast, the observed emission around the protostar position is weak and compact. The only way to reproduce this observational feature is with a wide-angle wind, for which electron density and ionization fraction would decrease much faster than in a collimated jet. Thus, we can easily reproduce the central compact source with a conical wind ($\epsilon$=1) with a velocity of 100~km~s$^{-1}$ and a mass-loss rate of $\dot{M}$=1.75$\times$10$^{-6}$ M$_\odot$~yr$^{-1}$ (see Figure \ref{Fig2}c). We also set the temperature and ionization fractions to decrease as a consequence of the fast decrease in density as the material moves away from the protostar (see Reynolds 1986). For a wind with these characteristics, at the angular resolution and sensitivity of our VLA observations, we are only detecting emission from the densest part at the very center of the flow (see Figure \ref{Fig2}c). The two elongated knots can be reproduced by simply adding to the conical wind a standard collimated jet but, this time, we assume the jet to start farther away from the protostar. Given the angular resolution of the observations (28 mas) and the jet inclination (26$^\circ$ with respect to the plane of the sky), the collimated jet should start at a distance $z_0\gtrsim$20~au in order to separate its emission from that of the central compact source. We found that we can reproduce reasonably well the observed morphology with the jet starting at distances up to $z_0\simeq$30~au. In Figure \ref{Fig3}, we show a model with the collimated jet starting at a distance of 25~au. We found that a mass-loss rate of $\dot{M}$=1.5$\times$10$^{-6}$ M$_\odot$~yr$^{-1}$ approximately reproduces the maximum intensity seen for the northern knot. Finally, the model matches most of the observational features by assuming that the P.A.s of the jet and counter-jet (part of the jet moving away from the observer) are slightly misaligned and that the southwestern knot has a lower electron density which could be explained by an intrinsically less dense counter-jet, and/or a lower ionization fraction, and/or a lower mass-loss rate (see Figures \ref{Fig2}e). Note that the inclination of the jet/disk system results in emission of the southern knot passing through the disk. Therefore, another possibility is that, even with similar physical parameters, the observed difference in brightness could be a consequence of some extinction of the emission from the southern knot when passing through very dense material in the disk (see Figure \ref{Fig3}).

\section{Discussion}

The presence of highly collimated jets associated with both low- and high-mass protostars is usually invoked as evidence of a similar formation mechanism for stars of all masses (e.g., Rosero et al. 2019, Purser et al. 2021). In the case of low-mass protostars, it has been well established that radio continuum jets appear elongated at distances smaller than 100~au from the protostar (see review by Anglada et al. 2018), and the highest angular resolution observations confirm that jets are already collimated even at distances of a few au (e.g., Carrasco-Gonz\'alez et al. 2019). These results are consistent with the expectations of both the X- and Disk-wind models, where collimation takes place at the scales of the protostar and disk. Studying the physical scales of collimation around high-mass protostars has been more difficult than for their low-mass counterparts, mainly because the former objects are usually found at much larger distances ($\gtrsim$1~kpc) and require a much higher resolution. Interestingly, as mentioned before, massive protostars have also been commonly found driving poorly collimated or even spherical winds (e.g., Torrelles et al. 1997, 2001, 2011; Moscadelli et al. 2007), which suggests that the whole process of mass ejection in these systems has some additional physics we are missing. Here, we observe into the innermost $\sim$100~au of a radio jet driven by a massive protostar for the first time, revealing a rather different morphology than that commonly observed in low-mass radio jets. We observe that the radio continuum near Cep A HW2 seems to be tracing two outflowing components (see Figure \ref{Fig3}): a poorly collimated wind launched from a distance of $\sim$1~au, and a highly collimated jet starting at distances of $\sim$20-30~au from the protostar. Moreover, we note that the mass-loss rates of the wide-angle wind and jet components are consistent with a constant flow of the order of 10$^{-6}$ M$_\odot$~yr$^{-1}$. Thus, our results point towards a scenario where the outflowing material is ejected very near the protostar over a wide angle (namely, a wind), and, farther away, it is collimated into a narrow stream (namely, a jet). We propose two scenarios that can possibly explain these results: 1) a disk-wind from a massive protostar/disk system, and 2) an externally collimated jet. Each scenario has very different consequences on the star formation process, as we discuss below.

 The first possibility is that the launching and collimation mechanism in this massive object is similar to that of low-mass jets, but the collimation distance simply scales with the mass of the protostar. This is plausible in the framework of a disk-wind, whose terminal wind/jet velocity ($v_{inf}$) is a function of the keplerian velocity ($v_{K,0}$) at the radius, $r_0$, where the wind is launched (Pudritz et al. 2007):

\begin{equation}
v_{inf} = \sqrt{2} \ v_{K,0} \ (r_A/r0) = \sqrt{2 G M_{star}/r_0} \ (r_A/r_0), 
\end{equation}

\noindent where $r_A$ is the radius at which the wind attains the Alfven velocity. In this model, the wind is initially launched from the disk with a wide opening angle, and collimates into a cylindrical jet at a distance from the protostar $z\gtrsim r_A \simeq 3 r_0$ (Pudritz et al. 2007). For a 10\,$M_\odot$ protostar, the radius where the wind is launched will be ten times larger than that of a 1 $M_\odot$ protostar at a fixed velocity. Therefore, it is possible that, while jets from low-mass protostars can be efficiently collimated at a few au from the protostar, the same mechanism requires some tens of au in order to collimate disk-winds from protostars much more massive. Future magneto-hydrodynamic simulations, aimed to extend the results of low-mass jets to higher masses, can test whether this scenario is possible. If that is the case, then our results could be interpreted as new evidence for an universal mechanism of jet collimation regulating the whole spectrum of stellar masses.

A second possibility is that, in order to achieve a high degree of collimation, jets from high-mass protostars might require restrictive physical conditions in the ambient medium, which will promote external collimation at large distances from the protostar. Recent works support a shock-ionization mechanism for jets from massive protostars, similar to what is observed in low-mass protostellar jets (e.g., Rosero et al. 2019, Fedriani et al. 2019, Purser et al. 2021). This result discards radiatively driven winds from massive protostars and supports an universal launching mechanism for stars of all masses (e.g., Purser et al. 2021). However, massive protostars are known to accrete at large rates from very massive and most probably unstable disks (e.g., Sanna et al. 2019b, Johnston et al. 2020). Under these conditions, it might be very difficult to sustain in time a strong helical magnetic field as required by the X- and Disk-wind models in order to collimate a jet in the vicinity of the protostar/disk system. In contrast, it has been proposed that wide-angle winds could be externally collimated into a jet by the presence of a large scale ordered magnetic field and/or a dense ambient medium (Albertazzi et al. 2014, Carrasco-Gonz\'alez et al. 2015). These physical conditions are indeed present in the vicinity of Cep A HW2, which is embedded in a high density molecular core (Torrelles et al. 1993, 2007) where an ordered magnetic field oriented parallel to the jet axis was detected both near the protostar (Vlemmings et al. 2010) and at larger scales (Curran \& Chrysostomou 2007). Therefore, an interesting possibility is that, in the case of a high-mass protostar, the launching mechanism could result in a poorly collimated wind, which could be collimated farther away into a jet under favorable ambient conditions only. When observed at larger scales, these jets would still look similar to their low-mass counterparts, although having different origins. In this case, our results would point to a possible difference in the nature of the ejection and collimation mechanism driven by protostars of different masses which, in turn, would imply different accretion histories due to the connection between outflow and accretion.

\noindent \emph{Acknowledgments.} We thank Henrik Beuther and Aina Palau for providing us the NOEMA 1.3 mm image. This work was supported by UNAM DGAPA-PAPIIT grants IN108218, IN104319 and IG101321 and CONACyT Ciencia de Frontera grant number 86372. JMT acknowledges partial support from the State Agency for Research (AEI) of the Spanish MCIU through the AYA2017-84390-C2 grant (cofunded with FEDER funds). Software: CASA (McMullin et al. 2007), \emph{sf3dmodels} (Izquierdo et al. 2018), RADMC-3D (Dullemond et al. 2012). AFI acknowledges support from the Deutsche Forschungs-gemeinschaft (DFG, German Research Foundation) - Ref no. FOR 2634/1 TE 1024/1-1. We thank an anonymous referee for a constructive review of the manuscript.


\end{document}